\documentclass[letterpaper,12pt]{article}

\usepackage[margin=1in]{geometry}

\usepackage{amsmath, amssymb}

\usepackage[hyphens]{url}
\usepackage{hyperref}

\usepackage{multirow}

\usepackage{csvsimple} 
\usepackage{booktabs}
\usepackage{longtable}  

\usepackage{graphicx}
\usepackage{float}
\usepackage{subcaption}

\usepackage{microtype}
\usepackage{seqsplit}
\newcommand{\filename}[1]{\texttt{\seqsplit{#1}}}

\title{NFL Draft Modelling: Loss Functional Analysis}
\author{
    \textbf{Tanmay Grandhisiri} \\
    Michigan State University \\
    \texttt{grandhi1@msu.edu}
}

\begin{document}

\maketitle

\begin{abstract}
In the NFL draft, teams must strategically balance immediate player impact against long-term value, presenting a complex optimization challenge for draft capital management. This paper introduces a framework for evaluating the fairness and efficiency of draft pick trades using norm-based loss functions. Draft pick valuations are modelled by the Weibull distribution. Utilizing these valuation techniques, the research identifies key trade-offs between aggressive, immediate-impact strategies and conservative, risk-averse approaches. Ultimately, this framework serves as a valuable analytical tool for assessing NFL draft trade fairness and value distribution, aiding team decision-makers and enriching insights within the sports analytics community.
\end{abstract}

\textbf{Reproducibility statement}: The code and the data in this analysis is reproducible and publicly available on Github: \url{https://github.com/tanmay-sketch/draftdynamics}


\section{Introduction}
The NFL draft is a yearly event where teams get the opportunity to strengthen their rosters with new players. There are 7 rounds in an NFL draft and each of the 32 clubs receives one pick in each of the seven rounds of the draft. The order of selection is determined by the reverse order of the finish in the previous season. This however is excluding any trades in between the clubs. Once the teams are assigned their draft positions, each pick now becomes an asset that the team can choose to trade. It's up to the team's executives to either select a player or trade the pick to another team to improve its possition in the current or future drafts. Teams may negotiate trades at any time before and during the draft and can swap picks or current NFL players to whom they hold the rights \cite{NFLDraftRules}. Turns out, this is a crucial part of the NFL Draft where teams are constantly evaluating their choices and it is the team's responsibility to optimize their trades and obtain the most value. This research introduces a framework for quantifying the fairness and efficiency of draft pick trades through norm-based loss functions. \\ 

Draft pick valuations are modeled using the exponential Massey-Thaler curve \cite{massey2013loserscurse}, with fitted parameters capturing market dynamics. The analysis explores a norm-based loss functions, $L^{1}$ and $L^{2}$ norms (Mean Absolute Error and Mean Squared Error), to evaluate how different strategic philosophies prioritize draft capital. By comparing the outcomes of various loss functions, fitted to NFL trades over the last 20 years, this study provides insights into the underlying trade-offs between aggressive, top-heavy strategies and more balanced, risk-averse approaches. The framework offers a flexible tool for assessing the fairness and value distribution of NFL draft trades, with potential applications for both team decision-making and broader sports analytics.\newline

\textbf{Why Norm based loss functions?} In traditional machine learning a loss function is used to calculate the error between the actual and predicted values of a model. In this analysis, we are using the concept of norms as a "strategy to draft" to fit a prior distribution that Massey and Thaler had proposed. The strategy with which a team chooses to go is implemented in the way in which we minimize the difference in value between the trades, similar to the concept of residuals. 

\section{Background Information}

From the Loser's curse paper \cite{massey2013loserscurse} we know that the relative value of subsequent picks can be given by the formula: 
\[
v_n(\alpha,\beta)=e^{-\lambda(n-1)^\beta}
\]

This is a two parameter distribution where we have to find the optimal pair of $(\alpha,\beta)$ that minimizes the error between the trades.This can be generalized to a multi-player draft pick swap, where $m$ picks are swapped for $n$ picks, and the summed values can be equated:
\[
\begin{aligned}
\mathbf{Value}(\text{team X's draft picks})
&= v_{j_1} + v_{j_2} + \dots + v_{j_m} \\
&\approx v_{k_1} + v_{k_2} + \dots + v_{k_n} \\
&= \mathbf{Value}(\text{team Y's draft picks}).
\end{aligned}
\]

Below is a formula that can be used to generalize the fairness in trades by ensuring that the aggregated value of one team’s traded picks matches that of the other. This approach builds upon established frameworks, as detailed in \cite{nathanson2023draft} (‘Exploring the Evolution of the NFL Draft Pick Trade Market Over Time’). By tuning $\lambda$,$\beta$ for different $p$ norms from actual trades, we can assess how different weighting schemes affect trade evaluations. 

\[
\begin{aligned}
\mathbf{Value}(\text{team X's draft picks})
&= \left( \sum_{i=1}^{m} \Bigl( e^{-\lambda (j_i-1)^\beta} \Bigr)^p \right)^{\frac{1}{p}}, \\
&\approx \left( \sum_{i=1}^{n} \Bigl( e^{-\lambda (k_i-1)^\beta} \Bigr)^p \right)^{\frac{1}{p}}, \\
&= \mathbf{Value}(\text{team Y's draft picks})
\end{aligned}
\]

\begin{itemize}
    \item \textbf{$\lambda$ (Lambda)}: Controls the rate at which draft pick value decreases. Higher $\lambda$ means quicker value drop-off.
    \item \textbf{$\beta$ (Beta)}: Adjusts the curvature of the decay in value. $\beta<1$ flattens the curve (later picks retain value), while $\beta>1$ steepens it (earlier picks more valuable).
    \item $p$ (Norm Parameter): Determines how individual pick values are combined into an overall measure, with common choices being $p=1$ and $p=2$ ($L^1$ and $L^2$ norm).
    \item \textbf{$j_i$}: The pick position that team X trades away.
    \item \textbf{$k_i$}: The pick position that team Y trades away.
\end{itemize}

In this optimization framework, we aim to minimize the error term defined by the difference between the aggregated values of the two teams’ draft picks. This error term represents the deviation between the calculated value for team X’s draft picks and that for team Y’s picks. By adjusting the parameters $\lambda$ and $\beta$, the goal is to minimize this discrepancy—typically through a least-squares or other norm-based approach—to ensure that both teams receive equivalent value in the trade. 

\[
\Delta = \bigg( \sum_{i=1}^{m} \big(e^{-\lambda (j_i-1)^\beta} \big)^p \bigg)^{\frac{1}{p}}
- \bigg( \sum_{i=1}^{n} \big(e^{-\lambda (k_i-1)^\beta} \big)^p \bigg)^{\frac{1}{p}}
\]

As mentioned before, fitting these $(\alpha,\beta)$ parameters require some kind of optimization. This is where norms are helpful, as they enable us to try different optimization strategies that align with a team’s focus on early, mid or late-round picks.\\

Both $L^{1}$ and $L^{2}$ norms are widely used in optimization and each has its own properties. 
The $L^{1}$ Norm follows the following formula: 
\[
\|x\|_1 = \sum_{i=1}^{n} |x_i|
\]

It is convex but not differentiable at $x_i = 0$ due to the absolute value function. This promotes sparsity in the solutions.

The $L^{2}$ Norm follows the following formula: 
\[
\|x\|_2 = \sqrt{\sum_{i=1}^{n} x_i^2}
\]

It is both convex and differentiable everywhere except at the origin. This differentiablity makes the $L^{2}$ norm particularly common for gradient based optimization methods. 

\begin{figure}[H]
    \centering
    \begin{subfigure}{0.24\textwidth}
        \centering
        \includegraphics[width=\textwidth]{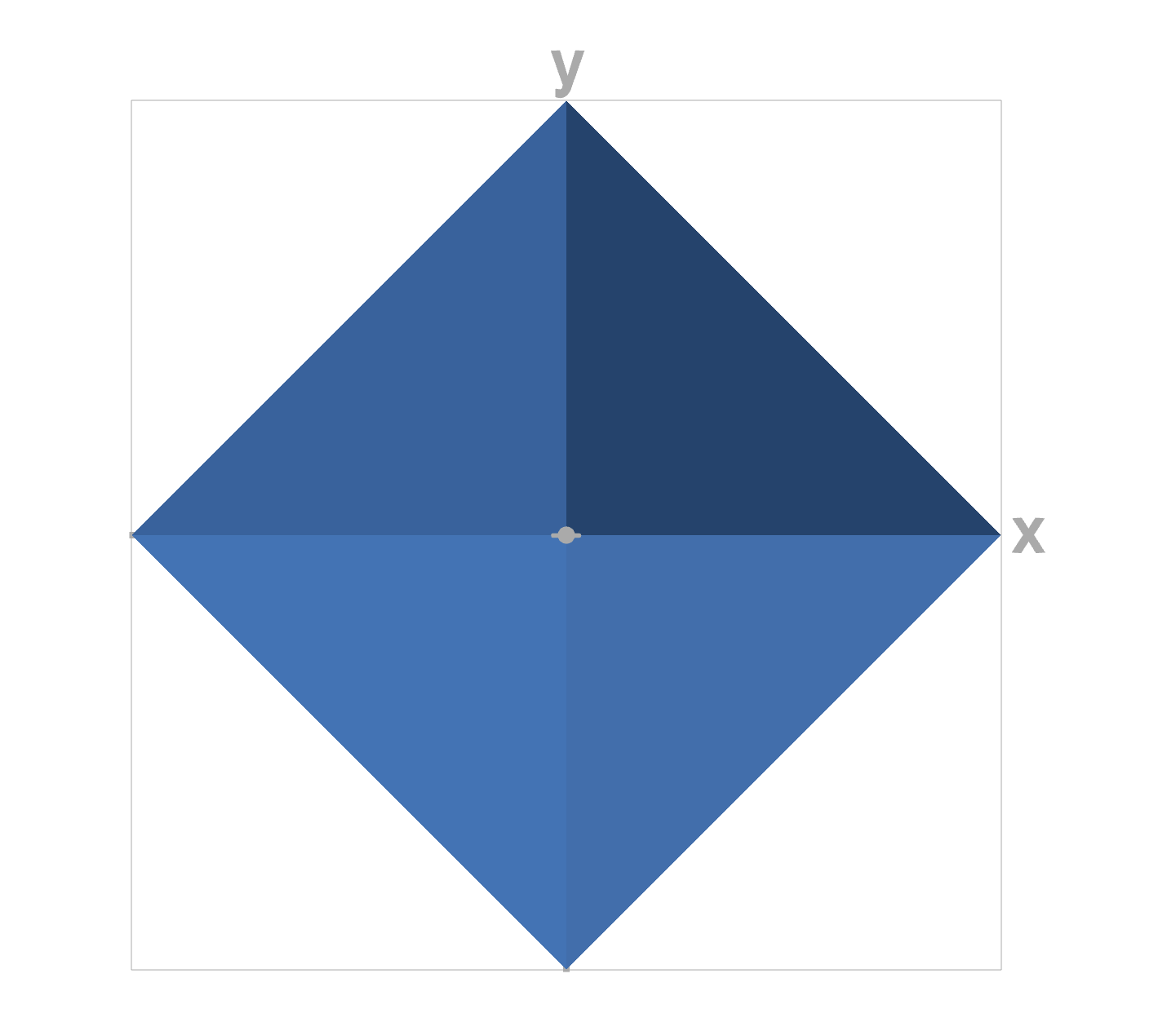}
        \caption*{L1 Norm (2D)}
        \label{fig:l1_norm_2d}
    \end{subfigure}
    \hfill
    \begin{subfigure}{0.24\textwidth}
        \centering
        \includegraphics[width=\textwidth]{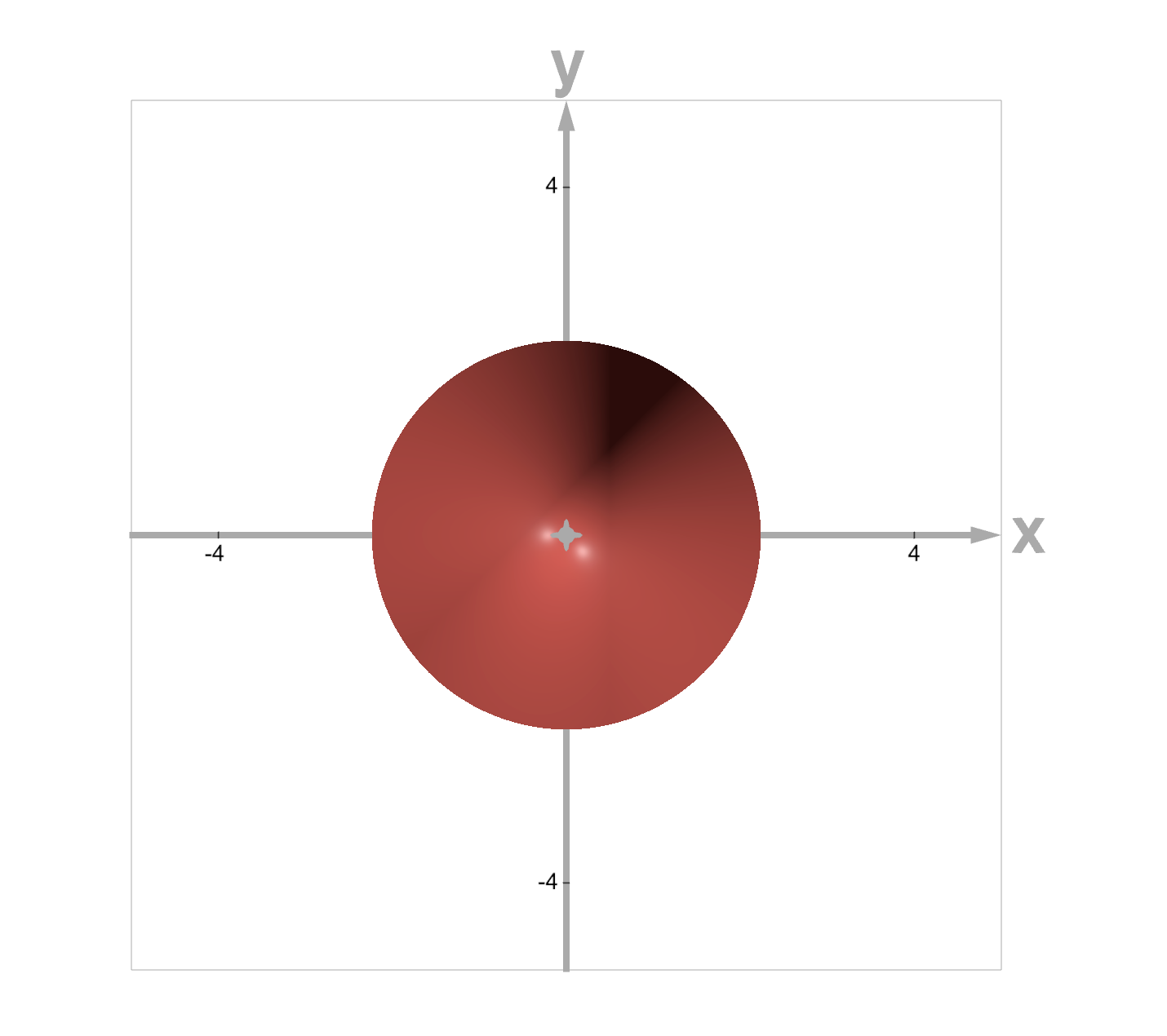}
        \caption*{L2 Norm (2D)}
        \label{fig:l2_norm_2d}
    \end{subfigure}
    \hfill
    \begin{subfigure}{0.24\textwidth}
        \centering
        \includegraphics[width=\textwidth]{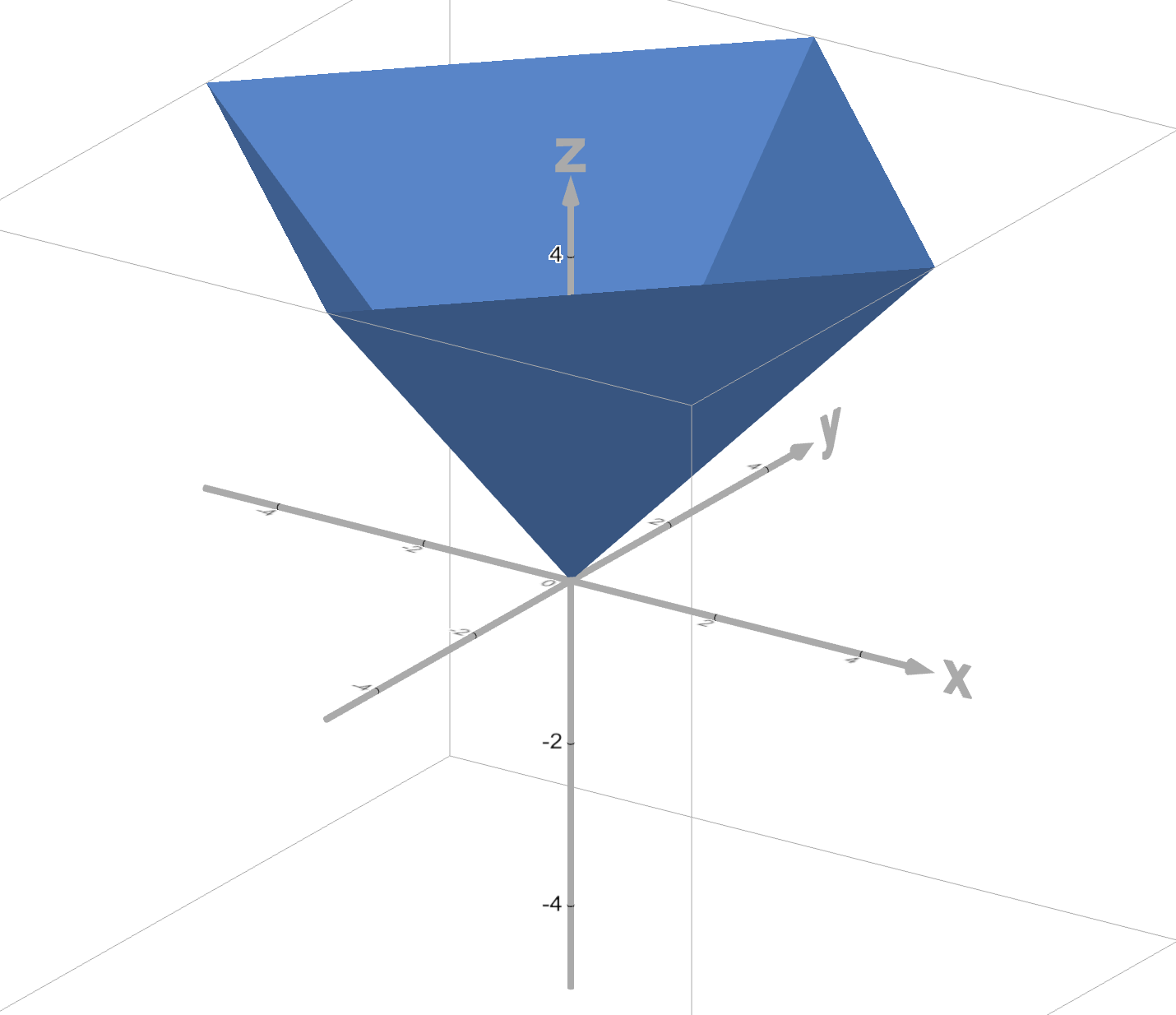}
        \caption*{L1 Norm (3D)}
        \label{fig:l1_norm}
    \end{subfigure}
    \hfill
    \begin{subfigure}{0.24\textwidth}
        \centering
        \includegraphics[width=\textwidth]{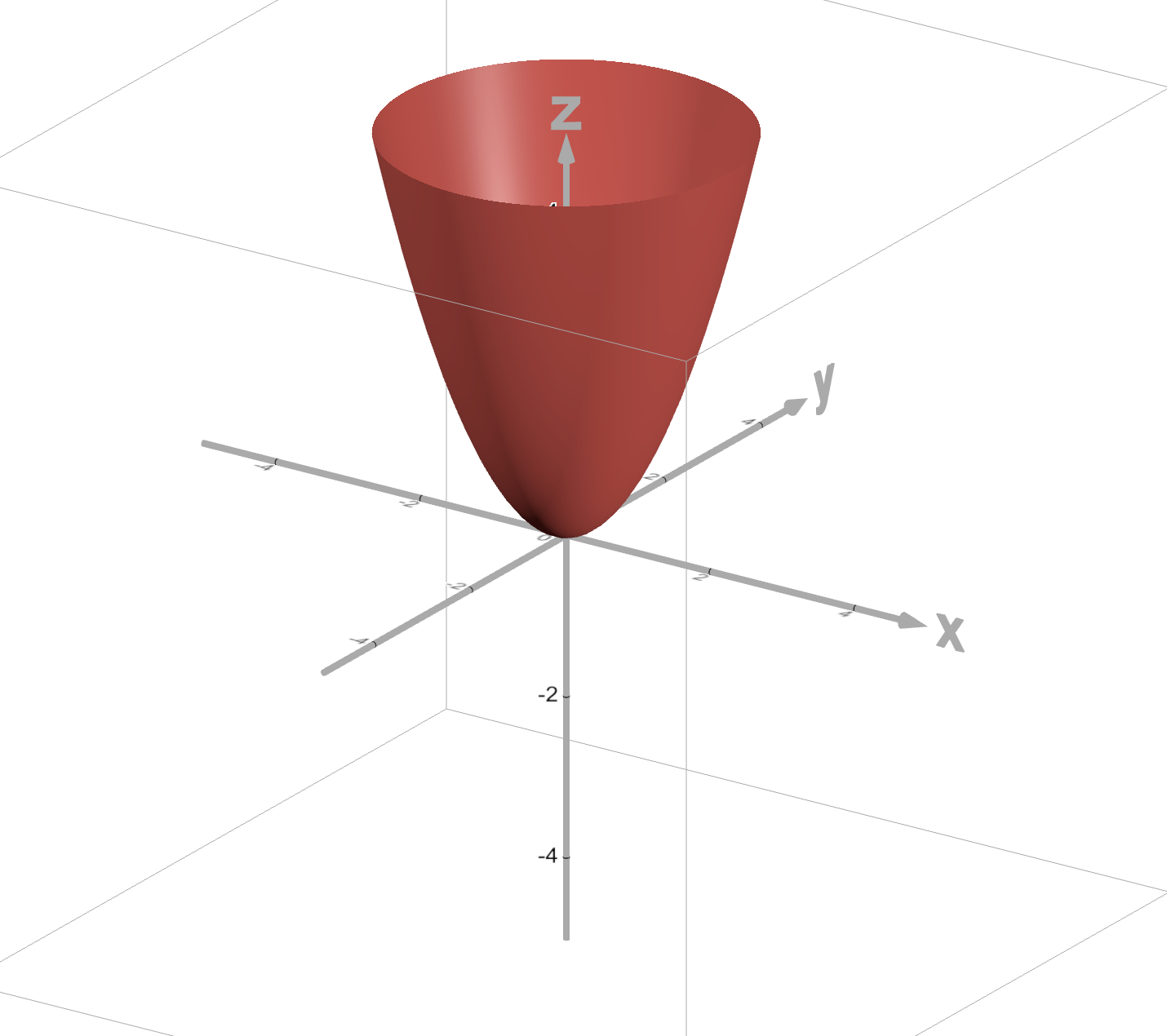}
        \caption*{L2 Norm (3D)}
        \label{fig:l2_norm}
    \end{subfigure}
    \caption{Comparison of L1 and L2 norms (2D and 3D).}
    \label{fig:l1_l2_comparison}
\end{figure}


\section{Methodology}

The analysis involves several critical stages, including preprocessing trade data, optimizing a trade valuation model, quantifying uncertainty through bootstrap resampling, and visualizing the findings. The data set used for this study is sourced from the \filename{data\_draft\_trades\_m24\_Chris.csv}, dataset which is publicly available on GitHub at \url{https://github.com/snoopryan123/NFL_draft_chart_Ryan}. The dataset is processed using Python's \texttt{pandas} library and is filtered to include trades from the time period 2006 to 2023. During the preprocessing phase, draft pick numbers are meticulously extracted by parsing columns that are initially formatted as strings. This parsing effectively segments the dataset into two distinct categories: upward draft picks, representing the selections traded away by teams aiming to move higher in the draft order, and downward draft picks, signifying the selections gained by teams opting to move lower. To enhance simplicity and consistency within the valuation framework, any future draft picks involved in these trades are explicitly excluded from consideration in this analysis, allowing for a clearer interpretation and valuation of the immediate trade impacts. While it is common for draft pick trades to include players and future picks or swaps, these are beyond the scope of the current analysis. \newline

Optimization of the trade valuation model involves fitting parameters using two convex and quasi-convex loss functions, Mean Squared Error (MSE) and Mean Absolute Error (MAE). Due to the convexity and differentiability properties of these loss functions, gradient-based methods are suitable, with the \textbf{L-BFGS-B} algorithm specifically chosen for its efficiency in handling optimization problems involving norms. Initial parameters of $[0.146, 0.698]$ were selected from preliminary analysis, and the optimization is capped at 1000 iterations to ensure convergence. \newline

To quantify uncertainty in parameter estimates, bootstrap resampling was utilized to compute the confidence intervals. This involves repeatedly sampling from the preprocessed data with replacement, re-estimating the parameters via the same optimization on each bootstrap sample, and constructing an empirical distribution of parameter estimates. Confidence intervals at the 95\% confidence level for parameters $\lambda$ and $\beta$ are then determined from the 2.5th and 97.5th percentiles of these bootstrap distributions.\newline

Finally, a visualization is created by plotting fitted valuation curves along with their bootstrap-derived 95\% confidence intervals, allowing for clear visual interpretation. 


\section{Analysis}

The results highlight key strategic differences in draft approaches—MSE optimization heavily favors early picks, reflecting a risk-averse strategy that prioritizes top talent, whereas MAE optimization distributes value more evenly across all selections, suggesting a greater emphasis on depth and long-term team development. By examining these valuation curves and their confidence intervals, we can better understand how teams assess draft capital and make trade decisions to maximize efficiency and competitive advantage. \newline \newline \newline

\begin{figure}[H]
    \centering
    \begin{subfigure}[b]{0.48\textwidth}
        \centering
        \includegraphics[width=\textwidth]{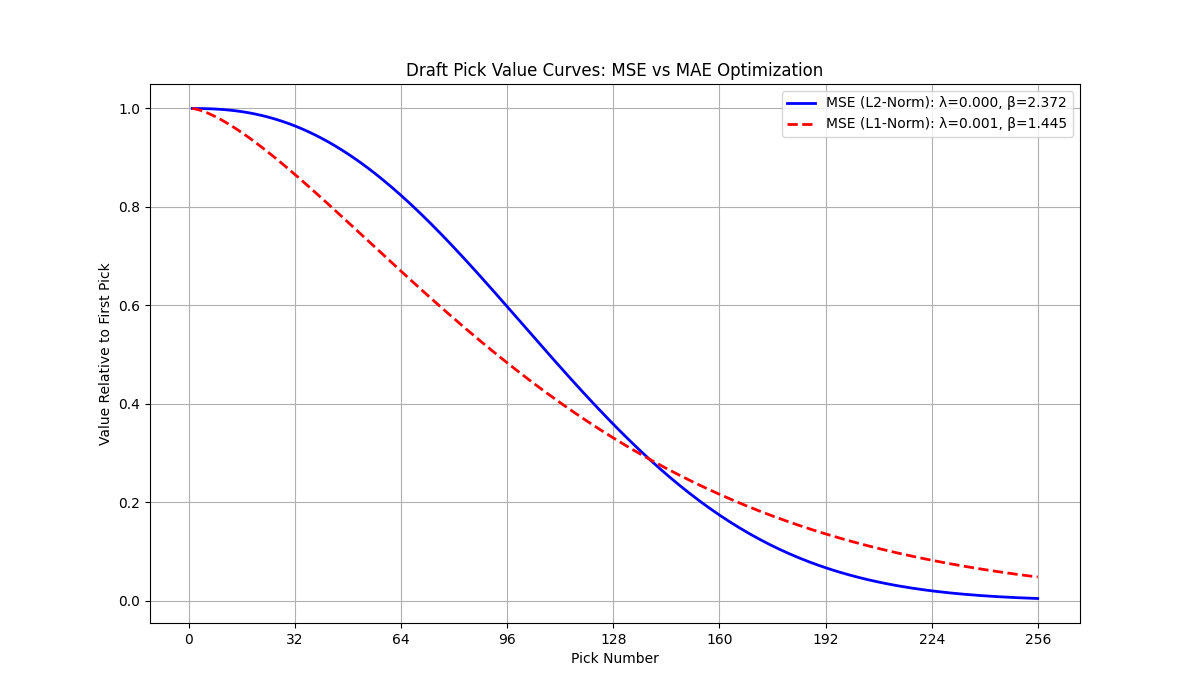}
        \caption{$L^1$ vs $L^2$ value curves}
        \label{fig:draft_value_curve}
    \end{subfigure}
    \hfill
    \begin{subfigure}[b]{0.48\textwidth}
        \centering
        \includegraphics[width=\textwidth]{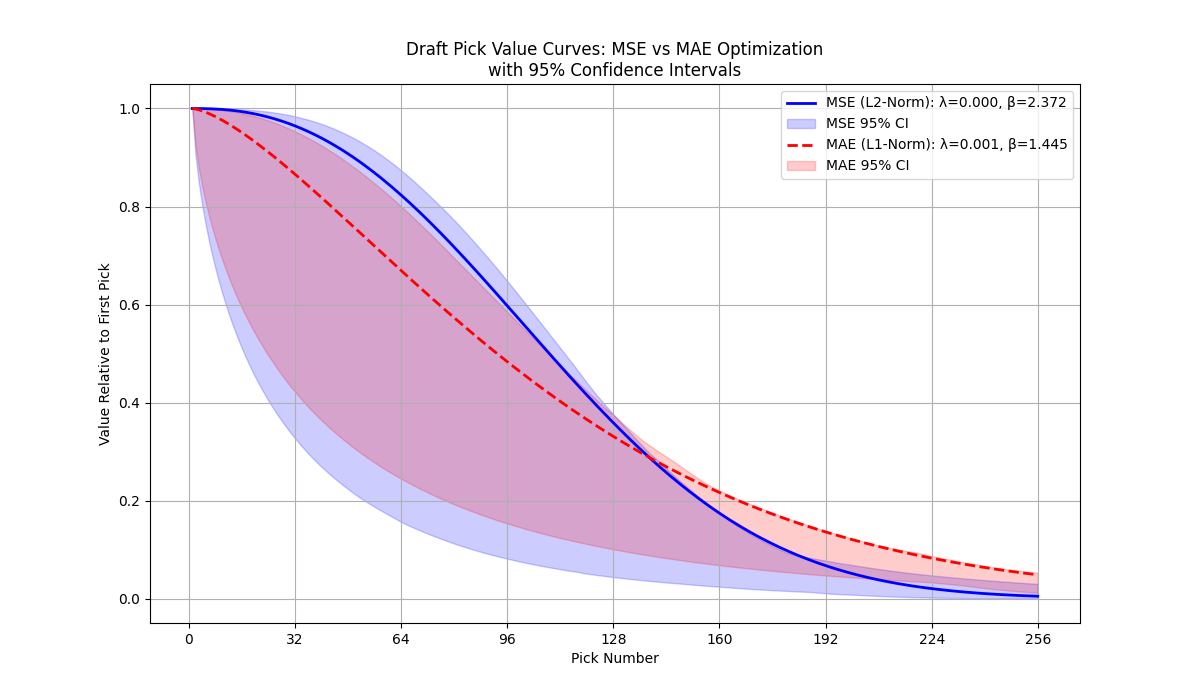}
        \caption{Value Curves with 95\% CIs}
        \label{fig:bootstrapped_samples}
    \end{subfigure}
    \caption{Draft value curves comparison}
    \label{fig:combined_figures}
\end{figure}

Figures \ref{fig:draft_value_curve} and \ref{fig:bootstrapped_samples} show how MSE and MAE optimizations affect draft pick valuations. The MSE curve (L2 norm) drops sharply for early picks, emphasizing top selections, while the MAE curve (L1 norm) declines more gradually, spreading value across all picks. \newline

After fitting the value curves, the following table gives the estimated $\beta$ and $\lambda$ values. 

\begin{table}[ht]
\centering
\begin{tabular}{|c|c|c|c|c|}
\hline
\multirow{2}{*}{\textbf{Norm}} 
 & \multicolumn{2}{c|}{\textbf{Without CI}} 
 & \multicolumn{2}{c|}{\textbf{With CI}} \\
\cline{2-5}
 & \(\lambda\) & \(\beta\) & \(\lambda\) & \(\beta\) \\
 \hline
\(L^1\) 
 & 0.001007 & 1.445350
 & 0.015761 & 1.157242 \\
\hline
\(L^2\) 
 & 0.000010 & 2.371845 
 & 0.030773 & 1.638750 \\
\hline
\end{tabular}
\caption{Parameter Estimates for \(L^2\) (MSE) and \(L^1\) (MAE) Optimizations}
\label{tab:params}
\end{table}

MAE optimization assigns higher value to later picks, suggesting a strategy that prioritizes depth over top-heavy talent. Conversely, MSE optimization supports investing heavily in earlier draft picks. \textbf{Why do we observe this?} This difference arises from how L1 and L2 norms handle errors. The L2 norm (MSE) penalizes large deviations more heavily due to its squared nature. This leads to a model that strongly favors minimizing large errors, which results in prioritizing high-value picks and diminishing the value of later picks significantly. On the other hand, the L1 norm (MAE) treats all errors linearly, allowing a more balanced distribution of pick values. This results in a valuation strategy where later picks retain more value relative to the first pick. \newline

Additionally, papers like the loser's curse in draft analysis have shown that while top picks often carry superstar potential, mid-to-late round picks can still yield significant value through team fit, development, and cost efficiency \cite{massey2013loserscurse}. The MAE-based approach aligns with this broader value distribution, whereas the MSE-based approach assumes that diminishing returns occur rapidly as draft positions increase. \newline

The choice between these optimization strategies reflects different team philosophies: franchises seeking to maximize immediate star power may prefer an MSE-based approach, while those focusing on long-term depth and player development may benefit more from an MAE-based valuation.The 95\% confidence intervals in Figure \ref{fig:bootstrapped_samples} highlight estimation reliability, with greater uncertainty in later picks, particularly in MAE-based models.\\

Based on the parameters in Table \ref{tab:params}, we can generate a chart similar to the highly utilized Jimmy Johnson draft value chart \cite{drafttekChart}. The values obtained from $L^1$ and $L^2$ optimization are rescaled to align with those in the Jimmy Johnson chart, as detailed in Table \ref{tab:compare_jj}. The plot below compares the draft value curves with the Jimmy Johnson values.

\begin{figure}[H]
    \centering
    \includegraphics[width=0.5\textwidth]{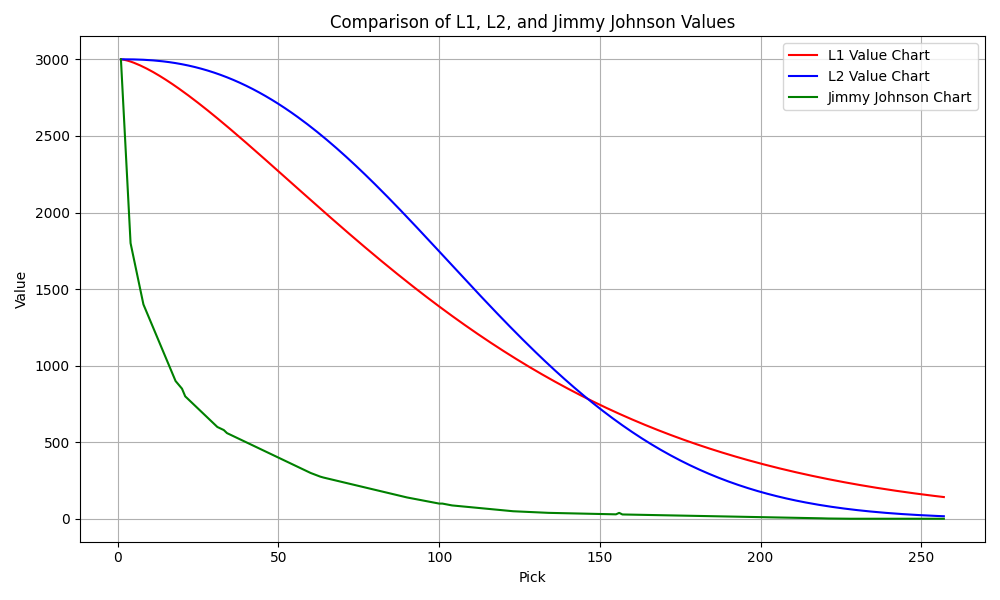}
    \caption{$L^1$, $L^2$, Jimmy Johnson Value Chart}
    \label{fig:jimmy_johnson_chart}
\end{figure}

The comparison of the $L^1$, $L^2$, and Jimmy Johnson draft value charts in Figure~\ref{fig:jimmy_johnson_chart} highlights distinct differences in draft pick valuation strategies. The Jimmy Johnson chart, originally created by the Dallas Cowboys in the early 1990s, was primarily a chart designed for quickly evaluating trade scenarios rather than being directly tied to empirical player outcomes \cite{jjValueChart}. Meanwhile, the $L^1$ and $L^2$ norms were mathematically optimized to minimize the difference in trade values, providing alternative views on pick valuation. Notably, while the $L^2$ norm closely resembles the Jimmy Johnson chart by steeply diminishing the value of later picks, the $L^1$ norm assigns considerably higher values to these same selections. This discrepancy arises because the $L^1$ norm minimizes absolute differences, inherently producing a more gradual decay in value, whereas the squared differences emphasized in the $L^2$ norm penalizes large deviations, resulting in a sharper drop-off. Practically, the relatively high valuation of late-round picks under the $L^1$ norm may not accurately reflect the real-world draft outcomes, as the probability of finding impactful players diminishes significantly towards the end of the draft. Therefore, while the $L^1$ norm can serve as a useful theoretical model for evaluating fairness in trades, it likely overstates the practical value of late-round picks compared to average late-round picks.


\section{Conclusion}
This study explored how using different valuation methods—Mean Squared Error (MSE) and Mean Absolute Error (MAE)—affects the way NFL teams value draft picks. The MSE method strongly favors early-round picks, making them seem far more valuable than later selections. On the other hand, the MAE method treats all picks more equally, giving greater value to picks in later rounds. These differences matter because they reflect different team-building strategies: teams using MSE are more likely to chase high-value players at the top of the draft, while those using MAE might build deeper teams by valuing later picks more highly. The direct exchange of draft picks is not often the case for trades between teams, especially picks in the same draft. Player's rights, cash, future picks are more common in trade deals in the modern NFL. Future studies could examine how these valuation methods connect these different assets and how actual player success can help teams improve their drafting strategy.

\nocite{*}
\bibliographystyle{plain}
\bibliography{main}

\newpage

\appendix
\section*{Appendix}
\renewcommand{\thesubsection}{\Alph{subsection}}

\subsection{Value Charts}

\begin{center}
    \begin{longtable}{cccc} 
      \toprule
      \textbf{Pick} & \textbf{L1\_value} & \textbf{L2\_value} & \textbf{Jimmy\_Johnson} \\
      \midrule
      \endfirsthead
      \toprule
      \textbf{Pick} & \textbf{L1\_value} & \textbf{L2\_value} & \textbf{Jimmy\_Johnson} \\
      \midrule
      \endhead
      \bottomrule
      \\
      \caption{Draft Comparison Data from \texttt{compare\_jj.csv}}%
      \label{tab:compare_jj}\\
      \endlastfoot
      \begin{filecontents*}{data/compare_jj.csv}
Pick,L1_Value,L2_Value,Jimmy_Johnson
1,3000.0,3000.0,3000
2,2996.98,2999.97,2600
3,2991.78,2999.84,2200
4,2985.25,2999.59,1800
5,2977.68,2999.2,1700
6,2969.23,2998.64,1600
7,2960.01,2997.9,1500
8,2950.11,2996.97,1400
9,2939.6,2995.84,1350
10,2928.53,2994.5,1300
11,2916.94,2992.95,1250
12,2904.86,2991.16,1200
13,2892.35,2989.14,1150
14,2879.41,2986.87,1100
15,2866.09,2984.35,1050
16,2852.39,2981.58,1000
17,2838.36,2978.54,950
18,2824.0,2975.24,900
19,2809.33,2971.66,875
20,2794.38,2967.81,850
21,2779.15,2963.67,800
22,2763.67,2959.24,780
23,2747.94,2954.52,760
24,2731.98,2949.51,740
25,2715.81,2944.19,720
26,2699.43,2938.58,700
27,2682.86,2932.66,680
28,2666.11,2926.43,660
29,2649.19,2919.89,640
30,2632.1,2913.04,620
31,2614.86,2905.87,600
32,2597.49,2898.39,590
33,2579.97,2890.59,580
34,2562.33,2882.47,560
35,2544.58,2874.03,550
36,2526.72,2865.27,540
37,2508.75,2856.18,530
38,2490.7,2846.78,520
39,2472.55,2837.05,510
40,2454.33,2827.0,500
41,2436.03,2816.63,490
42,2417.67,2805.93,480
43,2399.25,2794.92,470
44,2380.77,2783.58,460
45,2362.25,2771.92,450
46,2343.68,2759.95,440
47,2325.08,2747.66,430
48,2306.45,2735.06,420
49,2287.79,2722.14,410
50,2269.11,2708.91,400
51,2250.41,2695.37,390
52,2231.71,2681.53,380
53,2212.99,2667.38,370
54,2194.28,2652.93,360
55,2175.56,2638.19,350
56,2156.86,2623.15,340
57,2138.16,2607.82,330
58,2119.48,2592.2,320
59,2100.81,2576.3,310
60,2082.17,2560.12,300
61,2063.56,2543.67,292
62,2044.97,2526.94,284
63,2026.42,2509.95,276
64,2007.91,2492.69,270
65,1989.43,2475.18,265
66,1971.0,2457.41,260
67,1952.61,2439.4,255
68,1934.27,2421.15,250
69,1915.99,2402.66,245
70,1897.75,2383.93,240
71,1879.58,2364.98,235
72,1861.46,2345.82,230
73,1843.41,2326.43,225
74,1825.42,2306.84,220
75,1807.5,2287.05,215
76,1789.65,2267.06,210
77,1771.87,2246.89,205
78,1754.16,2226.53,200
79,1736.53,2205.99,195
80,1718.98,2185.28,190
81,1701.5,2164.41,185
82,1684.11,2143.39,180
83,1666.8,2122.21,175
84,1649.58,2100.89,170
85,1632.44,2079.44,165
86,1615.39,2057.85,160
87,1598.43,2036.15,155
88,1581.57,2014.33,150
89,1564.79,1992.4,145
90,1548.11,1970.37,140
91,1531.53,1948.25,136
92,1515.04,1926.05,132
93,1498.65,1903.76,128
94,1482.36,1881.41,124
95,1466.17,1858.99,120
96,1450.08,1836.51,116
97,1434.09,1813.99,112
98,1418.2,1791.42,108
99,1402.42,1768.82,104
100,1386.75,1746.19,100
101,1371.18,1723.54,100
102,1355.72,1700.88,96
103,1340.36,1678.21,92
104,1325.11,1655.55,88
105,1309.97,1632.89,86
106,1294.94,1610.25,84
107,1280.02,1587.63,82
108,1265.21,1565.03,80
109,1250.51,1542.48,78
110,1235.92,1519.97,76
111,1221.44,1497.51,74
112,1207.08,1475.1,72
113,1192.82,1452.76,70
114,1178.68,1430.48,68
115,1164.66,1408.28,66
116,1150.74,1386.17,64
117,1136.94,1364.14,62
118,1123.25,1342.21,60
119,1109.68,1320.37,58
120,1096.21,1298.65,56
121,1082.87,1277.03,54
122,1069.63,1255.53,52
123,1056.51,1234.16,50
124,1043.51,1212.91,49
125,1030.62,1191.8,48
126,1017.84,1170.82,47
127,1005.17,1149.99,46
128,992.62,1129.31,45
129,980.18,1108.78,44
130,967.86,1088.41,43
131,955.65,1068.2,42
132,943.55,1048.16,41
133,931.56,1028.29,40
134,919.69,1008.6,39.5
135,907.93,989.08,39
136,896.28,969.75,38.5
137,884.74,950.6,38
138,873.32,931.64,37.5
139,862.0,912.88,37
140,850.8,894.31,36.5
141,839.71,875.94,36
142,828.72,857.77,35.5
143,817.85,839.81,35
144,807.08,822.05,34.5
145,796.43,804.51,34
146,785.88,787.17,33.5
147,775.44,770.05,33
148,765.11,753.15,32.6
149,754.88,736.46,32.3
150,744.77,720.0,31.8
151,734.75,703.75,31.4
152,724.84,687.73,31
153,715.04,671.93,30.6
154,705.34,656.35,30.2
155,695.75,641.0,29.8
156,686.26,625.88,29.4
157,676.87,610.98,29
158,667.58,596.31,28.6
159,658.4,581.87,28.2
160,649.32,567.66,27.8
161,640.33,553.68,27.4
162,631.45,539.92,27
163,622.66,526.4,26.6
164,613.98,513.1,26.2
165,605.39,500.03,25.8
166,596.9,487.18,25.4
167,588.51,474.57,25
168,580.21,462.18,24.6
169,572.0,450.01,24.2
170,563.9,438.07,23.8
171,555.88,426.35,23.4
172,547.96,414.86,23
173,540.13,403.58,22.6
174,532.4,392.53,22.2
175,524.75,381.69,21.8
176,517.2,371.08,21.4
177,509.73,360.67,21
178,502.36,350.48,20.6
179,495.07,340.5,19.8
180,487.87,330.74,19.4
181,480.76,321.18,19
182,473.74,311.82,18.6
183,466.8,302.67,18.2
184,459.94,293.73,17.8
185,453.17,284.98,17.4
186,446.49,276.43,17
187,439.88,268.08,16.6
188,433.36,259.92,16.2
189,426.92,251.95,15.8
190,420.56,244.17,15.4
191,414.28,236.58,15
192,408.08,229.17,14.6
193,401.96,221.94,14.2
194,395.92,214.89,13.8
195,389.95,208.01,13.4
196,384.06,201.31,13
197,378.25,194.78,12.6
198,372.51,188.42,12.2
199,366.85,182.22,11.8
200,361.25,176.19,11.4
201,355.74,170.32,11
202,350.29,164.6,10.6
203,344.92,159.04,10.2
204,339.61,153.63,9.8
205,334.38,148.37,9.4
206,329.22,143.26,9
207,324.12,138.29,8.6
208,319.1,133.46,8.2
209,314.14,128.77,7.8
210,309.25,124.21,7.4
211,304.42,119.79,7
212,299.66,115.49,6.6
213,294.96,111.33,6.2
214,290.33,107.29,5.8
215,285.76,103.37,5.4
216,281.25,99.57,5
217,276.81,95.89,4.6
218,272.42,92.32,4.2
219,268.1,88.86,3.8
220,263.84,85.51,3
221,259.63,82.27,2.6
222,255.49,79.13,2.3
223,251.4,76.09,2
224,247.37,73.16,1.8
225,243.4,70.31,1.6
226,239.48,67.57,1.4
227,235.62,64.91,1.2
228,231.82,62.34,1
229,228.06,59.86,1
230,224.37,57.47,1
231,220.72,55.15,1
232,217.13,52.92,1
233,213.58,50.76,1
234,210.09,48.69,1
235,206.65,46.68,1
236,203.26,44.75,1
237,199.92,42.88,1
238,196.63,41.08,1
239,193.39,39.35,1
240,190.19,37.68,1
241,187.04,36.08,1
242,183.94,34.53,1
243,180.88,33.04,1
244,177.87,31.61,1
245,174.9,30.23,1
246,171.98,28.91,1
247,169.1,27.63,1
248,166.27,26.41,1
249,163.47,25.23,1
250,160.72,24.1,1
251,158.01,23.02,1
252,155.34,21.97,1
253,152.72,20.97,1
254,150.13,20.01,1
255,147.58,19.09,1
256,145.07,18.21,1
257,142.6,17.36,1
\end{filecontents*}
\csvreader[
        separator=comma,
        late after line=\\,
        respect underscore
      ]{data/compare_jj.csv}{}%
      {\csvcoli & \csvcolii & \csvcoliii & \csvcoliv}
    \end{longtable}
\end{center}

\end{document}